\documentclass{article}

\usepackage{arxiv}

\usepackage[utf8]{inputenc} 
\usepackage[T1]{fontenc}    
\usepackage{hyperref}       
\usepackage{url}            
\usepackage{booktabs}       
\usepackage{amsfonts}       
\usepackage{nicefrac}       
\usepackage{microtype}      
\usepackage{lipsum}
\usepackage{graphicx}
\usepackage{amsmath}
\usepackage{amsfonts}
\usepackage{amssymb}
\usepackage{subfigure}
\usepackage{multirow}
\usepackage{float}
\usepackage{adjustbox}
\usepackage{booktabs}
\usepackage{lscape}
\usepackage{enumitem, array}

\title{RECURRENT-TYPE NEURAL NETWORKS FOR REAL-TIME SHORT-TERM PREDICTION OF SHIP MOTIONS IN HIGH SEA STATE}

\author{
 Danny D'Agostino$^{1,2}$, Andrea Serani$^{2,\star}$, Frederick Stern$^3$, and Matteo Diez$^{2}$ \\
 \\
  $^1$Department of Computer, Control, and Management Engineering ``A. Ruberti'', Sapienza University of Rome, Italy\\
  $^2$CNR-INM, National Research Council-Institute of Marine Engineering, Italy\\
  $^3$IIHR-Hydroscience \& Engineering, The University of Iowa, Iowa City, USA\\
  $^\star$Corresponding author: \texttt{andrea.serani@cnr.it} \\
}

\begin{document}
\maketitle

\begin{abstract}
The prediction capability of recurrent-type neural networks is investigated for real-time short-term prediction (nowcasting) of ship motions in high sea state. Specifically, the performance of recurrent neural networks, long-short term memory, and gated recurrent units models are assessed and compared using a data set coming from computational fluid dynamics simulations of a self-propelled destroyer-type vessel in stern-quartering sea state 7. Time series of incident wave, ship motions, rudder angle, as well as immersion probes, are used as variables for a nowcasting problem. The objective is to obtain about 20 s ahead prediction. Overall, the three methods provide promising and comparable results.
\end{abstract}


\section{Introduction}
The prediction of the seakeeping and maneuverability performance of naval ships constitutes one of the most challenging problems in naval hydrodynamics and is important from both an operational and safety point of views, especially in heavy weather conditions.
Seakeeping and maneuverability of naval ships in heavy weather has been traditionally investigated by means of experimental model scale testing in large basins. 
In order to reduce the statistical uncertainty of the experimental campaigns and to met security and safety as for the NATO Standardization Agreement, a large number of conditions (i.e., speeds, wave headings, length, and height, number of encounters wave) have to be investigated during the tests, including so-called rare events. This makes scale model testing time consuming and expensive.

During the last decades low- to high-fidelity simulation methods have been developed for investigating ships seakeeping and maneuvering. Nevertheless, a complete solution of the seakeeping and maneuverability problem involves resolving complex nonlinear wave-body interactions that may require hundreds of computational CPU hours, especially if statistical indicators are sought after. 
For this reasons, to alleviate the computational burden associated with numerical simulations, machine learning methods, such as neural networks (NNs), can be used to model and predict seakeeping and maneuverability performance of ships. 

Classical NNs treat each observation or data point in the same way. This means that the NN does not take into account the correlation across the data points, assuming that they are independent and identically distributed (\textit{i.i.d.}). Nevertheless, in several application, such as in time series fore- and nowcasting (long- and real-time short-term predictions), the value of the target variable (e.g., ship motions and controllers) is usually strongly correlated to the past values of the target variable at the previous time step. This correlation is lost in a classical NN model. In order to solve this limitation, recurrent NNs (RNNs) have been developed with the objective to learn the dependencies of the data across time and to improve the prediction accuracy in case of sequential data \cite{rumelhart1986-N}. A RNN is a class of artificial neural networks where connections between nodes form a directed graph along a temporal sequence, allowing to exhibit temporal dynamic behavior. Derived from feedforward neural networks, RNNs can use their internal state (memory) to process inputs sequences of variable length. Nevertheless, RNN suffers the so called \textit{vanishing gradient problem} \cite{pascanu2013difficulty}. To overcame this issue, different mathematical models have been developed creating \textit{gates} along the time steps. Among them the long-short term memory (LSTM) \cite{hochreiter1997long} and the gated recurrent unit (GRU) \cite{cho2014learning} have shown quite effective performance for modeling sequences in several research fields. 

In the ship hydrodynamics context, the development and the assessment of machine learning methods in fore- and nowcasting of ship motions and (possibly) loads have become of certain interest and a cutting-edge topic in the ocean engineering community. In particular, recurrent-type NNs nowcasting capabilities results to be an hot topic of research. Trained by both historical and computational fluid dynamic (CFD) data, up to real-time data, NNs could provide decision support to captains in choosing route, heading, and speed, contributing to the safety of vessels, cargo, and crews.
Short-term prediction based on radial basis NN has been presented in \cite{de2011ship}. LSTM and GRU have been investigated for the prediction of 2 and 3 degrees of freedom (DoF) of a catamaran in sea state 1 and the DTMB model in sea state 8, based on CFD computations in \cite{del2021learning}. 

The objective of the present work is to assess the capability of recurrent-type NNs for real-time short-term prediction (nowcasting) of ship motions in high sea state. Specifically the performance of RNN, LSTM, and GRU models are assessed for the nowcasting of a self-propelled destroyer-type vessel, sailing in stern-quartering sea state 7.

The data set is formed by free-running CFD simulations of a destroyer-type vessel with appendages (skeg, twin split bilge keels, twin rudders and rudder seats slanted outwards, shafts, and struts), that have been assessed for course keeping in irregular stern-quartering waves (sea state 7) at target Froude number equal to 0.33, within the activity of the NATO STO Research Task Group AVT-280 ``Evaluation of Prediction Methods for Ship Performance in Heavy Weather'' \cite{van2020prediction}. RNN, LSTM, and GRU are assessed and compared in predicting wave elevation, ship motions, rudder angle, and immersion probes time histories.

\begin{figure}[!b]
\centering
\includegraphics[width=0.95\textwidth]{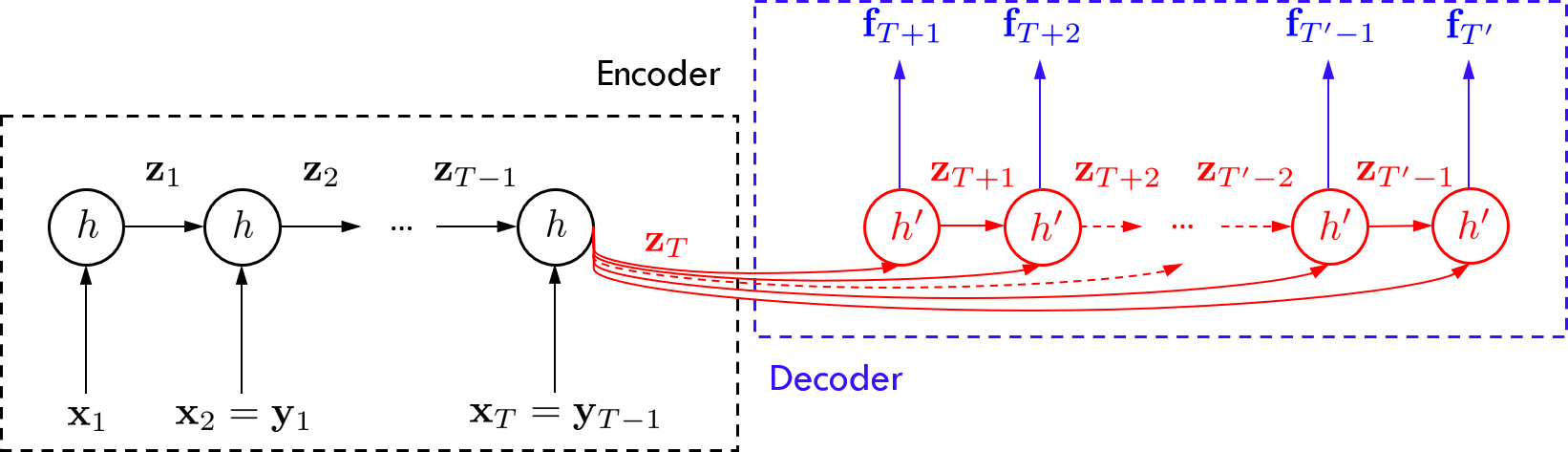}
\caption{Conceptualization of the sequence to sequence learning via an encoder-decoder model.}\label{fig:seqTOseq}
\end{figure}
\section{Recurrent-type NNs for Sequences Modeling}
A recurrent-type NN differs from a classical NN, allowing to pass at the successive time step the hidden units $\mathbf{z}_t$ or \textit{states} of the network as a function  of the input data $\mathbf{x}_t\in \mathbb{R}^D$ and the state at the previous time step $\mathbf{z}_{t-1}$, namely $\mathbf{z}_{t+1} = h(\mathbf{x}_t, \mathbf{z}_{t-1})$.

For fore- and nowcasting of time series data (or sequences modeling), observing the input data $\mathbf{x}_t\in \mathbb{R}^D$ for a temporal window $T$ ($t=1,\dots,T$), at $t=T$ a recurrent-type NNs can predict (in real time) multiple time steps $t'$ with $t'=T+1,\dots,T'$ of the target variable $\mathbf{y}\in\mathbb{R}^K$, with $T'$ non necessarily equal to $T$ (i.e. the length of the desired output may differs from the length of the input). This particular problem is called \textit{sequence to sequence} learning where the model is trained to map an input sequence of fixed length $\mathbf{x}_t$ for $t = 1, \dots, T$ which best predicts the target variables $\mathbf{y}_{t}$ for $t = T+1, \dots, T'$. A particular architecture that allows to model this kind of problems is the \textit{encoder-decoder} model developed for machine translation \cite{sutskever2014sequence}. The model is composed by two parts as shown in Figure \ref{fig:seqTOseq}: the encoder network which take all the inputs vector $\mathbf{x}_1, \dots, \mathbf{x}_{T}$ and return a latent representation of what the encoder learned in the time window $T$ namely the final hidden state $\mathbf{z}_{T}$ for $t = 1, \dots, T$, through the function $h(\mathbf{x}_t,\mathbf{z}_{t-1})$.
Given the vector $\mathbf{z}_{T}$, the decoder network will map into the target space $\mathbb{R}^K$ the latent representations for $t = T+1, \dots, T'$, through a function $\mathbf{z}_{t}=h'(\mathbf{z}_{T},\mathbf{z}_{t-1})$, providing 
\begin{align}
	\mathbf{f}_{t}  &= \mathbf{W}_{zf}\mathbf{z}_{t}
\end{align}
where $\mathbf{f}_t\in\mathbb{R}^K$ is the prediction of $\mathbf{y}_t$ and $\mathbf{W}_{zf}$ is a weights matrix of dimension $K\times M$, with $M$ an hyperparameter.

Network's hyperparameters are found minimizing the reconstruction error for the target, defined as follows
\begin{equation}\label{eq:mse1}
	e(\mathbf{W}) = \frac{1}{(T' - T)} \sum_{t=T}^{T'} ||\mathbf{y}_t - \mathbf{f}(\mathbf{z}_t, \mathbf{W})||^2
\end{equation}
Note that the NNs works with variables normalized within $-1$ and $1$.

\subsection{Recurrent Neural Networks}
The equations for the forward propagation of an RNN (for $t=1,\dots, T$) reads 
\begin{align}
	\mathbf{z}_{t} &= \tanh(\mathbf{W}_{xz}\mathbf{x}_{t} + \mathbf{W}_{zz}\mathbf{z}_{t-1})
\end{align}\label{eq:RNN}
with $T$ the time window and also the number of RNN's cells, $\tanh$ the hyperbolic tangent function applied element wise, $\mathbf{W}_{xz}$ and $\mathbf{W}_{zz}$ the weights matrices with dimension $M\times D$ and $M\times M$, respectively. Equation \ref{eq:RNN} is used for the encoding phase, while for the decoding $\mathbf{x}_t$ is substituted by $\mathbf{z}_T$.

\subsection{Long-Short Term Memory}
The LSTM cell or unit is composed by three main gates called the \textit{input} $\mathbf{i}$, \textit{forget} $\mathbf{g}$, \textit{output} $\mathbf{o}$, and the \textit{cell-state} $\mathbf{c}_t$. They are all $M$-dimensional vectors that cover a particular role in the network. Those are given by
\begin{equation}\label{eq:lstm}
	\begin{bmatrix}
		\mathbf{i} \\
		\mathbf{g} \\
		\mathbf{o} \\
		\mathbf{c} 
	\end{bmatrix}
	=
	\begin{bmatrix}
		\text{sigm} \\
		\text{sigm} \\
		\text{sigm} \\
		\text{tanh} 
	\end{bmatrix}
	\mathbf{W} 
	\begin{bmatrix}
		\mathbf{x}_t \\
		\mathbf{z}_t
	\end{bmatrix} 
\end{equation}
where sigm is the sigmoid function and the weights matrix $\mathbf{W}$ is of dimension $4M \times (M+D)$. The update of the cell state $\mathbf{c}_t$ and the state $\mathbf{z}_t$ is given by
\begin{align}
	\mathbf{c}_t &= \mathbf{g} \odot \mathbf{c}_{t-1} + \mathbf{i}\odot\mathbf{c} \\
	\mathbf{z}_t &= \mathbf{o} \odot \tanh(\mathbf{c}_t)
\end{align}
with ``$\odot$'' the Hadamard product. The vector $\mathbf{g}$ is called forget  because it multiplies by the cell state at the previous time step $\mathbf{c}_{t-1}$. Since $\mathbf{g}$ assume values between $0$ and $1$, this can be interpreted as the amount of information that are allowed to pass to the next cell state. The intermediate cell state vector $\mathbf{c}$ is multiplied by the input vector $\mathbf{i}$, which can be seen as what kind of new information could be relevant for the current cell state update. Finally, the state vector $\mathbf{z}_t$ is updated filtering the cell state vector $\mathbf{c}_t$ with a multiplication respect to output gate $\mathbf{o}$. Equation \ref{eq:lstm} is used for the encoding phase, while for the decoding $\mathbf{x}_t$ is substituted by $\mathbf{z}_T$.

\subsection{Gated Recurrent Units}
The mathematical model describing the state updates of a GRU is similar to the LSTM network, but it has only two gates as follows  
\begin{equation}\label{eq:GRU}
	\begin{bmatrix}
		\mathbf{d} \\
		\mathbf{r}  
	\end{bmatrix}
	=
	\begin{bmatrix}
		\text{sigm} \\
		\text{sigm} 
	\end{bmatrix}
	\mathbf{W}_1 
	\begin{bmatrix}
		\mathbf{x}_t \\
		\mathbf{z}_t
	\end{bmatrix} 
\end{equation}
where $\mathbf{d}$ and $\mathbf{r}$ are the update and the reset gates, respectively. The weights matrix $\mathbf{W}_1$ has dimension $2M \times (M+D)$. The state $\mathbf{z}_t$ update is given by
\begin{equation}
	\mathbf{z}_t = \mathbf{d}\odot\mathbf{z}_{t-1} + (1-\mathbf{d})\odot\tanh\bigg( \mathbf{W}_2\begin{bmatrix}
		\mathbf{x}_t \\
		\mathbf{r}\odot\mathbf{z}_{t-1}  
	\end{bmatrix}\bigg)
\end{equation}
where the weights matrix $\mathbf{W}_2$ has dimension $M \times (M+D)$, with $M$ the dimensionality $\mathbf{d}$ and $\mathbf{r}$. It can be observed the the reset gate decide which information should be retained from the previous hidden state $\mathbf{z}_{t-1}$. Equation \ref{eq:GRU} is used for the encoding phase, while for the decoding $\mathbf{x}_t$ is substituted by $\mathbf{z}_T$.

\section{Application for Ship Motion Nowcasting}
The hull form under investigation is the MARIN model 7967 which is equivalent to 5415M, used as test case for the NATO STO Research Task Group AVT-280 ``Evaluation of Prediction Methods for Ship Performance in Heavy Weather'' \cite{van2020prediction}. This is a geosim replica of the DTMB 5415 model with different appendages designed by MARIN. The DTMB 5415 is an open-to-public naval combatant hull geometry. The model was self-propelled and kept on course by a proportional-derivative (PD) controller actuating the rudders angle. 
\begin{figure}[!t]
\centering
\includegraphics[width=0.95\textwidth]{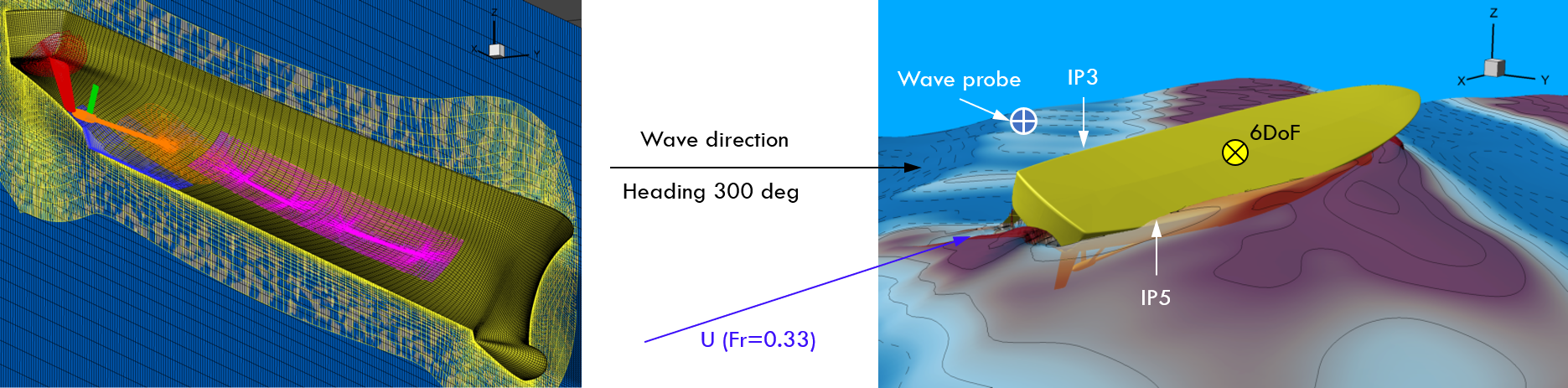}
\caption{Detail of the boundary-layer computational grid (left) and a CFD snapshot with location of the probes (right).}\label{fig:5415}
\end{figure}

The code CFDShip-Iowa V4.5 \cite{huang2008-IJNMF} is used for the CFD computations. CFDShip-Iowa is an overset, block structured CFD solver designed for ship applications using either an absolute or a relative inertial nonorthogonal curvilinear coordinate system for arbitrary moving but non-deforming control volumes. The free-running CFD simulations were performed with propeller RPM fixed to the self-propulsion point of the model for the envisaged speed. The simulations were conducted in irregular long crested waves, following a JONSWAP spectrum. The turbulence is computed by the isotropic Menter's blended $k-\epsilon/k-\omega$ (BKW) model with shear stress transport (SST) using no wall function. The location of the free surface is given by the ''zero'' value of the level-set function, positive in the water and negative in the air. The 6 degrees of freedom rigid body equations of motion are solved to calculate linear and angular motions of the ship. A simplified body-force model is used for the propeller, which prescribes axisymmetric body force with axial and tangential components. The total number of grid points is about 45M. Further details can be found in \cite{serani2021-OE}. 

The data set collects 8 CFD runs (with different random phases) at Fr = 0.33, with nominal peak period $T_p = 9.2$ s and wave heading of 300 deg. It may be noted that the simulation conditions are close to a resonance condition for the roll. The nominal significant wave height is equal to 7 m, corresponding to sea state 7 (high), according to the World Meteorological Organization (WMO) definition. A total of 215 encounter waves have been recorded, with a total run length of about 3323 s and a data rate equal to 129.2 Hz (for the current application the data set has been down-sampled to 8.6 Hz). Wave elevation far from the ship, ship motions (the 6 DoF), rudder angle, and two immersion probes (IP3 and IP5) time series compose the data set. Figure \ref{fig:5415} shows a detail of the computational grid (on left) and a snapshot of the ship behavior with the location of signal probes (on right). 

The main objective is to obtain an accurate real-time short-term prediction of about 20 s (about one and an half roll periods) of the ten variables ($D=10$) at the same time.

\section{Networks' Setup and Evaluation Metrics}
The dataset has been divided in $70\%$ training set, $24\%$ validation set, and $6\%$ test set, for cross-validation. 
The networks' hyperparameters are selected using a grid search by evaluating different layers (depth of the network, 1 and 2), number of hidden units $M$ (20, 50, 100, and 200), dropout percentage (0.1, 0.2, and 0.5). For the current analysis the batch size is fixed to 512 and the number of cells of the encoder/decoder network (width of the network) are fixed to 25 and 30 time steps, respectively, corresponding to about 18 s of observation in order to produce approximately 20 s of ahead prediction. The optimization is carried out using the Adam algorithm \cite{kingma2014adam} for a maximum number of epochs fixed to 1000. The early stopping strategy \cite{yao2007early} is used as regularizer.
A linear activation is used to compute the output vector $\mathbf{f}_{t}$. The same setting of the matrices parameters are used in each time step despite the states that can evolve in time. This \textit{parameter sharing} characteristic allows the network to generalize better even in case of limited number of training data \cite{goodfellow2016deep}. Furthermore, to improve the generalization, 200 Monte Carlo realization of the dropout are performed, providing the expected value and the variance (Var) of the prediction \cite{gal2016dropout,gal2016theoretically}. 
In the following, for the sake of simplicity, the prediction refers to the expected value, while the variance of the prediction is used to define the prediction uncertainty band as $\pm2\sqrt{\rm Var}$. 

Defining the network's residual (or error) at each time step $t$ for each variable (or feature) $i$ as follows
\begin{equation}
r_{i,t} = \frac{y_{i,t}-f_{i,t}}{2\sigma(y_i)},
\end{equation}
with $\sigma$ the signal standard deviation, the assessment of the network's performance are based on the evaluation of the normalized root mean squared error (NRMSE)
\begin{equation}
\mathrm{NRMSE}_i = \sqrt{\frac{1}{(T'- T)} \sum_{t=T+1}^{T'} r_{i,t}^2} \qquad \mathrm{with} \qquad i=1,\dots,D,
\end{equation}
as well as by evaluating the probability density functions (PDFs), via kernel density estimate (KDE), of the residuals and their statistical moments (i.e., mean, variance, skewness, and kurtosis). 

\begin{table}[b!]
\centering
\caption{Summary of the hyperparameters optimal set found via cross-validation.}
\label{tab:optimal_hp}
\begin{tabular}{cccccc}
\toprule
Model&  $M$ (encoder, decoder)  & Batch Size & Dropout & N. cells (encoder) & N. layers \\
\midrule
RNN  & (100, 100)  &512 & 0.2 & 25 & 1\\
GRU  & (100, 100)  &512 & 0.2 & 25 & 1\\
LSTM & (100, 100)  &512 & 0.2 & 25 & 1\\
\bottomrule
\end{tabular}
\end{table}

\begin{table}[!b]
\centering
\caption{NRMSE breakdown for each variable for training and test sets.}
\label{tab:results_var}
\begin{tabular}{ccccccc}
\hline
Data set & \multicolumn{3}{c}{Training} & \multicolumn{3}{c}{Test}\\
\midrule
Variable & RNN & GRU & LSTM  & RNN & GRU & LSTM \\
\midrule
Wave     & 0.079 & 0.057 & 0.052 & 0.186 & 0.175 & 0.191 \\
Surge    & 0.008 & 0.005 & 0.005 & 0.028 & 0.022 & 0.029 \\
Sway     & 0.022 & 0.012 & 0.011 & 0.059 & 0.075 & 0.115 \\
Heave    & 0.086 & 0.050 & 0.049 & 0.140 & 0.132 & 0.140 \\
Roll     & 0.017 & 0.010 & 0.009 & 0.026 & 0.025 & 0.026 \\
Pitch    & 0.062 & 0.036 & 0.035 & 0.121 & 0.105 & 0.102 \\
Yaw      & 0.046 & 0.031 & 0.028 & 0.146 & 0.135 & 0.151 \\
Rudder   & 0.024 & 0.016 & 0.014 & 0.037 & 0.033 & 0.036 \\
Im (IP3) & 0.068 & 0.041 & 0.038 & 0.111 & 0.106 & 0.121 \\
Im (IP5) & 0.100 & 0.062 & 0.057 & 0.156 & 0.146 & 0.154 \\
\midrule
Average  & 0.051 & 0.032 & 0.030 & 0.101 & 0.095 & 0.107 \\
\bottomrule
\end{tabular}
\end{table}
\section{Results and Discussion}
The optimal hyperparameters are given in Table \ref{tab:optimal_hp}. Interestingly, the three methods provide their optimal performance with the same hyperparameters (at least considering the current sets for the present application).

\begin{figure}[!t]
\centering
\includegraphics[width=1.\textwidth]{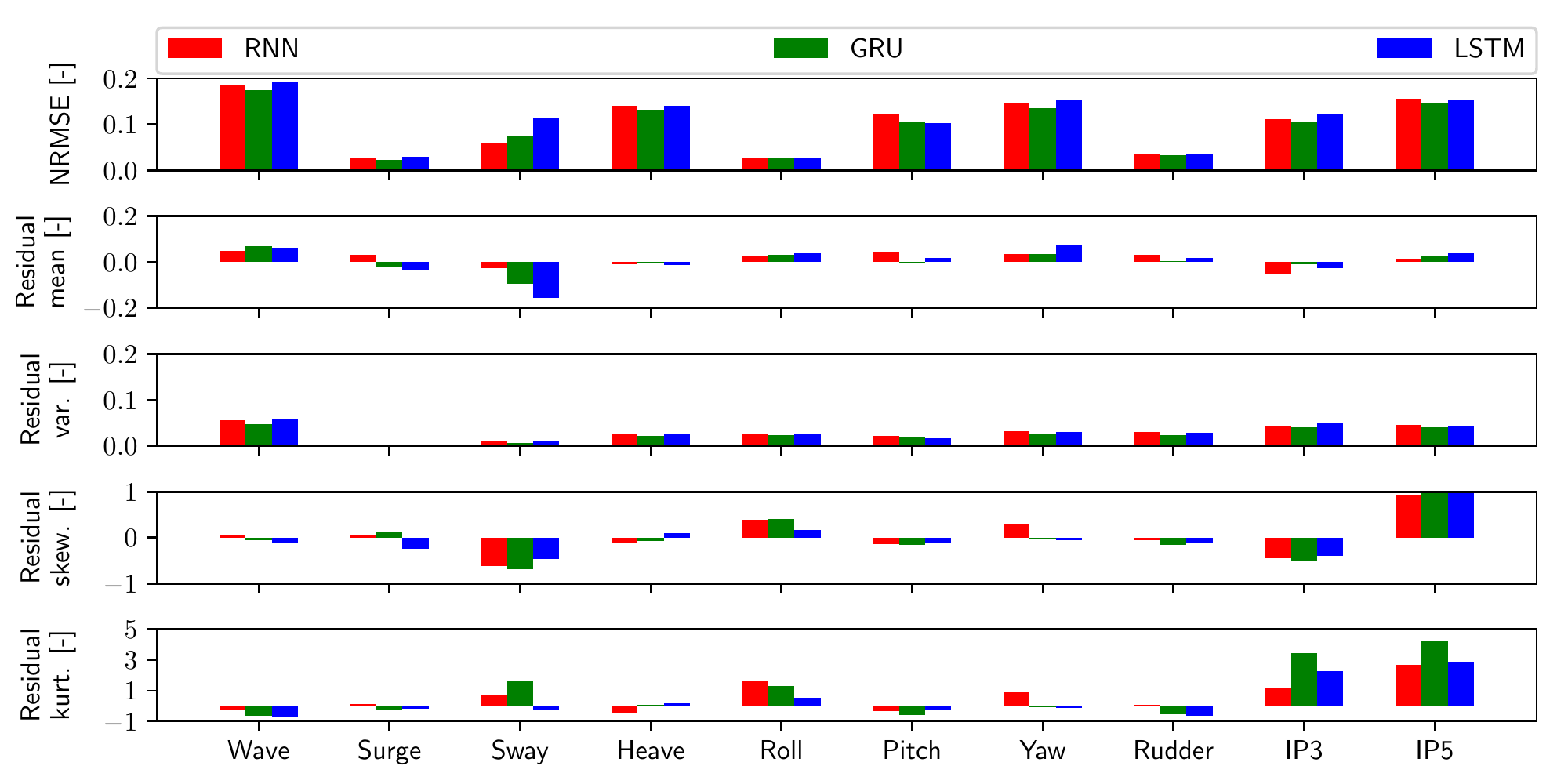}
\caption{Models NRMSE and statistical moments for the variables residuals.}\label{fig:kde_stats}
\end{figure}
\begin{figure}[!t]
\centering
\includegraphics[width=1.\textwidth]{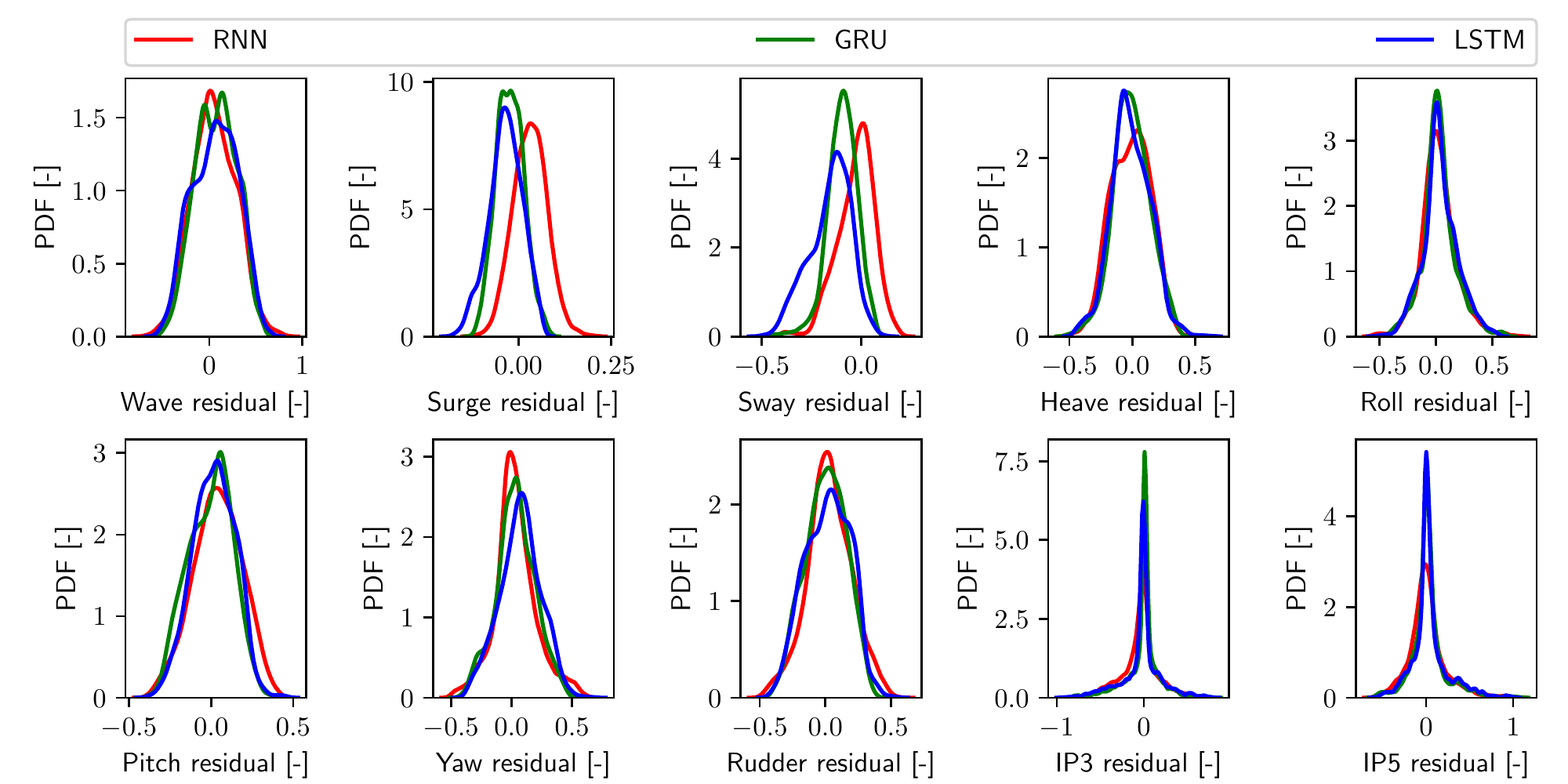}
\caption{PDF of the variables residuals by KDE.}\label{fig:kde_output}
\end{figure}
\begin{figure}[!t]
\centering
\includegraphics[width=1.\textwidth]{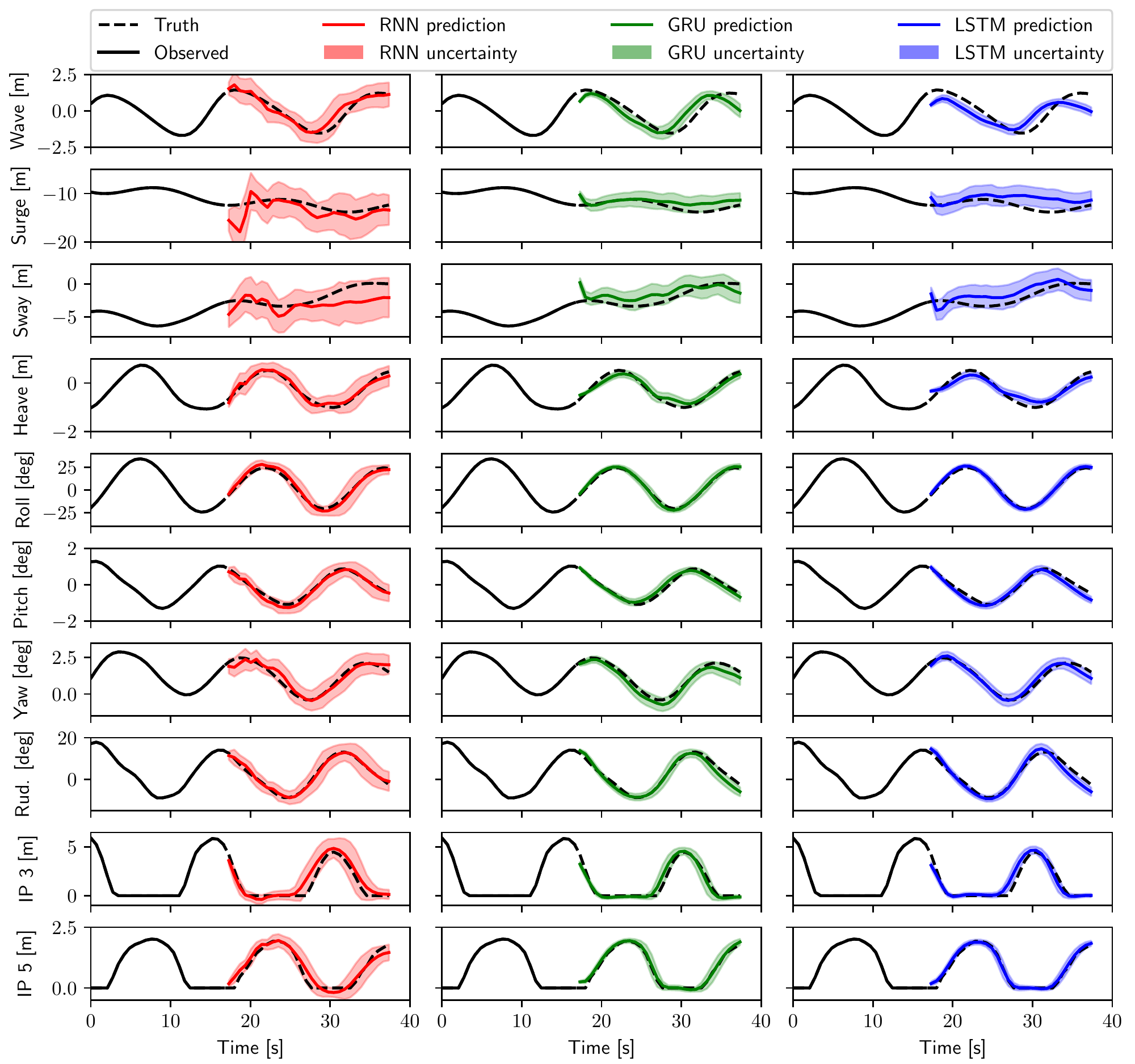}
\caption{Example of nowcasting for wave, ship motions, rudder angle, and immersion probes.}
\label{fig:WP2_f}
\end{figure}

Table \ref{tab:results_var} provides the average NRMSE for the training and the test sets obtained by each model. Furthermore, Table \ref{tab:results_var} shows the NRMSE for each variable, as well as Figure \ref{fig:kde_stats} (top row) for the test set. The lowest NRMSE on average for the test set is achieved by GRU followed by the RNN and LSTM models. The lowest NRMSE is achieved for surge, roll, and rudder angle. On the contrary, wave, heave, yaw, and immersion probe signals (IP3 and IP5) are the most challenging variables to nowcast, providing the highest errors. Overall, the performance of all the model are comparable, except for sway, where LSTM achieved the highest NRMSE with respect to the other models.

The PDFs of the variable residuals are shown in Figure \ref{fig:kde_output}. An important property that the residuals obtained from a fore- or nowcasting model should satisfy is that they should have a zero mean. In case of residuals with a mean strongly different from zero, it means that there is bias in the prediction and the model needs to be improved. Looking at Figure \ref{fig:kde_output}, the sway provides a mean slightly different from zero, especially for GRU and LSTM, while the RNN seems more robust in this case. Residuals mean values, as well as variance, skewness, and kurtosis are also shown in Fig. \ref{fig:kde_stats}. Wave has the highest variance. An high positive skewness (more weight in the right tail of the distribution) is obtained for the residuals of IP5 indicating a systematic overestimate of the forecast obtained for this variable, while the opposite behavior is obtained for sway and IP3. A substantial high value of the residuals kurtosis is obtained for both the immersion probes (IP3 and IP5), meaning that the distributions have long tails indicating the presence of high and low values in the residuals, as also shown in Figure \ref{fig:kde_output}. This is probably mainly due to the presence for IP3 and IP5 of strong changes from zero to higher values in some particular time step  which seems difficult to be modeled (high absolute value of the residuals), while for the rest of the time steps their values are very regular and simple to be predicted (low value of the residuals). 

Finally, an example of prediction expectation along with uncertainty band for each variable by all methods is shown in Figure \ref{fig:WP2_f}. An overall good prediction is achieved, even if some discrepancy is shown, specially for wave and sway, confirming the outcomes of the NMRSE and residuals assessment. It can also be observed that the wave considered is not the one acting on the ship's  center of gravity, but is the signal of a lateral probe (which provides a signal not affected by the ship's wake, see Figure \ref{fig:5415}). This means that between the processed wave and the ship system outputs (the 6 DoF, the rudder angle, and the immersion probes) there is a time lag, which ``relaxes'' the input/output relationship on the ship system state. For this reason it is possible that the NNs makes an higher error on the wave prediction.

\section{Conclusions and Future Work}
The performance of three recurrent-type neural networks for ship motion nowcasting have been assessed on a data set composed by CFD simulation of a self-propelled destroyer-type vessel in long-crest stern quartering waves at sea state 7.
Specifically, recurrent neural network, long-short term memory, and gated recurrent units have been assessed and compared for real-time short-term prediction of wave elevation, ship motions, rudder angle, and immersion probes time series. All the variables have been used defining a multiple time series nowcasting problem. The objective was to obtain about 20 s ahead prediction. 

An overall good performance is obtained using all the three methods. Surge, roll, and rudder angle prediction have provided the lowest errors, while wave and the immersion probes have shown the highest residuals. Overall, the GRU model have provided the best results, even if the three models are almost comparable. 

Future work will includes the use of the Bayesian optimization for the selection of the networks hyperparameters, as well as the analysis of the performance of the methods for real-time long-term prediction. Furthermore, to improve knowledge and forecasting of motions and trajectories for ships operating in waves, as well as global/local loads, hybrid machine learning methods will be also investigated \cite{diez2021-MARINE,diez2022-SNH}.

\section*{Acknowledgments}

CNR-INM is grateful to Drs. Elena McCarthy and Woei-Min Lin of the Office of Naval Research for their support through the Naval International Cooperative Opportunities in Science and Technology Program. Dr. Andrea Serani is also grateful to the National Research Council of Italy, for its support through the Short-Term Mobility Program 2018. The data set comes from the activity conducted within the NATO STO Research Task Group AVT-280 ``Evaluation of Prediction Methods for Ship Performance in Heavy Weather.''

\bibliographystyle{unsrt}  
\bibliography{biblio}  

\begin{thebibliography}{10}

\bibitem{rumelhart1986-N}
David~E Rumelhart, Geoffrey~E Hinton, and Ronald~J Williams.
\newblock Learning representations by back-propagating errors.
\newblock {\em Nature}, 323:533--536, 1986.

\bibitem{pascanu2013difficulty}
Razvan Pascanu, Tomas Mikolov, and Yoshua Bengio.
\newblock On the difficulty of training recurrent neural networks.
\newblock In {\em International conference on machine learning}, pages
  1310--1318, 2013.

\bibitem{hochreiter1997long}
Sepp Hochreiter and J{\"u}rgen Schmidhuber.
\newblock Long short-term memory.
\newblock {\em Neural computation}, 9(8):1735--1780, 1997.

\bibitem{cho2014learning}
Kyunghyun Cho, Bart Van~Merri{\"e}nboer, Caglar Gulcehre, Dzmitry Bahdanau,
  Fethi Bougares, Holger Schwenk, and Yoshua Bengio.
\newblock Learning phrase representations using rnn encoder-decoder for
  statistical machine translation.
\newblock {\em arXiv preprint arXiv:1406.1078}, 2014.

\bibitem{de2011ship}
Giulia De~Masi, Federico Gaggiotti, Roberto Bruschi, and Marco Venturi.
\newblock Ship motion prediction by radial basis neural networks.
\newblock In {\em 2011 IEEE Workshop On Hybrid Intelligent Models And
  Applications}, pages 28--32. IEEE, 2011.

\bibitem{del2021learning}
Jos{\'e} del {\'A}guila~Ferrandis, Michael~S Triantafyllou, Chryssostomos
  Chryssostomidis, and George~Em Karniadakis.
\newblock Learning functionals via {LSTM} neural networks for predicting vessel
  dynamics in extreme sea states.
\newblock {\em Proceedings of the Royal Society A}, 477(2245):20190897, 2021.

\bibitem{van2020prediction}
Frans van Walree, Andrea Serani, Matteo Diez, and Frederick Stern.
\newblock Prediction of heavy weather seakeeping of a destroyer hull form by
  means of time domain panel and cfd codes.
\newblock In {\em Proceedings of the 33nd Symposium on Naval Hydrodynamics,
  Osaka, Japan}, 2020.

\bibitem{sutskever2014sequence}
Ilya Sutskever, Oriol Vinyals, and Quoc~V Le.
\newblock Sequence to sequence learning with neural networks.
\newblock {\em Advances in neural information processing systems},
  27:3104--3112, 2014.

\bibitem{huang2008-IJNMF}
Juntao Huang, Pablo~M. Carrica, and Frederick Stern.
\newblock Semi-coupled air/water immersed boundary approach for curvilinear
  dynamic overset grids with application to ship hydrodynamics.
\newblock {\em International Journal for Numerical Methods in Fluids},
  58(6):591--624, 2008.

\bibitem{serani2021-OE}
A~Serani, M~Diez, F~van Walree, and F~Stern.
\newblock {URANS} simulations of a free-running destroyer sailing in irregular
  stern-quartering waves at sea state 7.
\newblock {\em Ocean Engineering}, 2021.
\newblock under review.

\bibitem{kingma2014adam}
Diederik~P Kingma and Jimmy Ba.
\newblock Adam: A method for stochastic optimization.
\newblock {\em arXiv preprint arXiv:1412.6980}, 2014.

\bibitem{yao2007early}
Yuan Yao, Lorenzo Rosasco, and Andrea Caponnetto.
\newblock On early stopping in gradient descent learning.
\newblock {\em Constructive Approximation}, 26(2):289--315, 2007.

\bibitem{goodfellow2016deep}
Ian Goodfellow, Yoshua Bengio, Aaron Courville, and Yoshua Bengio.
\newblock {\em Deep learning}.
\newblock MIT press Cambridge, 2016.

\bibitem{gal2016dropout}
Yarin Gal and Zoubin Ghahramani.
\newblock Dropout as a bayesian approximation: Representing model uncertainty
  in deep learning.
\newblock In {\em international conference on machine learning}, pages
  1050--1059, 2016.

\bibitem{gal2016theoretically}
Yarin Gal and Zoubin Ghahramani.
\newblock A theoretically grounded application of dropout in recurrent neural
  networks.
\newblock {\em Advances in neural information processing systems},
  29:1019--1027, 2016.

\bibitem{diez2021-MARINE}
M.~Diez, A.~Serani, E.~F. Campana, and F.~Stern.
\newblock Data-driven modeling of ship maneuvers in waves via dynamic mode
  decomposition.
\newblock In {\em Proceedings of the 9th International Conference on
  Computational Methods in Marine Engineering (Marine 2021)}, 2021.

\bibitem{diez2022-SNH}
M.~Diez, A.~Serani, M.~Gaggero, and E.~F. Campana.
\newblock Improving knowledge and forecasting of ship performance in waves via
  hybrid machine learning methods.
\newblock In {\em Proceedings of the 34th Symposium on Naval Hydrodynamics,
  Washington DC, USA}, 2022.

\end{thebibliography}

\end{document}